\documentclass[a4paper,11pt]{article}
\usepackage{contribution}

\newcommand{\weblink}[2][]{%
    \ifthenelse{\equal{#1}{}}%
    {\textnormal{\url{#2}}}%
    {\textnormal{\href{#2}{#1}}}%
}

\def\beq{\begin{equation}}
\def\eeq#1{\label{#1}\end{equation}}
\def\eeqn{\end{equation}}

\def\beqa{\begin{eqnarray}}
\def\eeqa#1{\label{#1}\end{eqnarray}}
\def\eeqan{\end{eqnarray}}

\let\bar=\overbar

\def\Dslash{\not{\hbox{\kern-4pt $D$}}}
\def\dslash{\not{\hbox{\kern-2pt $\del$}}}

\def\msb{{\bar{\ssstyle M \kern -1pt S}}}

\newcommand{\contribution}[7][]{%
  \clearpage
  \thispagestyle{plain}
  \ifthenelse{\equal{#1}{}}
  {\hypersetup{pdftitle={#2}}}
  {\hypersetup{pdftitle={#1}}}
  \hypersetup{pdfauthor={{#3} {#4}}}
  {\centering\normalfont\LARGE\bfseries\sffamily #2 \par\nobreak}
  \lhead{}
  \chead{%
    \textit{\footnotesize XIV International Conference on Hadron Spectroscopy
      (\weblink[\textit{hadron2011}]{http://www.hadron2011.de}), 13-17 June 2011, Munich, Germany}%
  }
  \rhead{}
  \bigskip
  \begin{center}
    {#3} {#4}\ifthenelse{\equal{#6}{}}{}{\footnote{\weblink[#6]{mailto:#6}}}
    \ifthenelse{\equal{#7}{}}{}{#7} \\
    \textit{#5}
  \end{center}
  \bigskip
}

\renewcommand{\abstract}[1]{%
  \begin{center}
    \begin{minipage}{0.85\textwidth}
      \begin{footnotesize}
        #1
      \end{footnotesize}
    \end{minipage}
  \end{center}
  \bigskip
}

\begin{document}
{\makeatletter\@ifundefined{c@affiliation}
{\newcounter{affiliation}}{}\makeatother
\newcommand{\affiliation}[2][]{\setcounter{affiliation}{#2}
\ensuremath{{^{\alph{affiliation}}}\text{#1}}}
\contribution[Heavy-Meson Decay Constants from Borel Sum Rules in
QCD]{Heavy-Quark Masses and Heavy-Meson Decay Constants from Borel
Sum Rules in QCD}{Dmitri}{Melikhov}{\affiliation[HEPHY, Austrian
Academy of Sciences, Nikolsdorfergasse 18, A-1050 Vienna,
Austria]{1}\\ \affiliation[Faculty of Physics, University of
Vienna, Boltzmanngasse 5, A-1090 Vienna, Austria]{2}\\
\affiliation[SINP, Moscow State University, 119991 Moscow,
Russia]{3}\\ \affiliation[INFN, Sezione di Roma Tre, Via della
Vasca Navale 84, I-00146 Roma, Italy]{4}}{dmitri\_melikhov@gmx.de}
{\!\!$^,\affiliation{1}^,\affiliation{2}^,\affiliation{3}$,
Wolfgang Lucha\!\affiliation{1}, and Silvano
Simula\!\affiliation{4}}

\abstract{Slight sophistications of the QCD sum-rule formalism may
have great impact on the reliability of predicted hadron
observables, as exemplified for the case of heavy-meson
decay~constants.}

{\bf Quark--Hadron Duality.} The extraction of the decay constant
$f_P$ of any ground-state heavy pseudoscalar meson $P$ from QCD
sum rules \cite{svz,lms2010,lms2011} is a two-phase process:
First, the operator product expansion (OPE) for the
Borel-transformed correlation function of the two relevant
pseudoscalar heavy-light currents has to be derived. Second, even
if all parameters of this OPE were known exactly, the knowledge of
merely {\em truncated\/} OPEs for correlators allows to extract
bound-state features with only a limited accuracy, reflecting an
intrinsic uncertainty of the QCD sum-rule formalism. Controlling
this uncertainty poses a delicate challenge \cite{lms_1}. We
consider mesons $P\equiv(Q\,\bar q)$ of mass $M_P$ composed of
heavy quarks $Q$ and light quarks~$\bar q.$ The {\em assumption\/}
of {\em quark--hadron duality\/} entails a relation between the
hadronic ground-state contribution and the QCD correlator
truncated at a certain {\em effective continuum threshold\/}
$s_{\rm eff}$:
\begin{equation}\label{SR_QCD}f_P^2\,M_P^4\exp(-M_P^2\,\tau)
=\Pi_{\rm dual}(\tau,s_{\rm eff})\equiv\int^{s_{\rm
eff}}_{(m_Q+m_q)^2}{\rm d}s\exp(-s\,\tau)\,\rho_{\rm pert}(s)
+\Pi_{\rm power}(\tau)\ .\end{equation}Obviously, in order to be
able to extract $f_P$ one has to develop a procedure determining
$s_{\rm eff}$. Borel transformations introduce a mass parameter
$\widetilde M,$ included here in the form $\tau\equiv1/\widetilde
M^2.$ A crucial, albeit rather trivial, observation is that
$s_{\rm eff}$ must be a function of $\tau.$ Otherwise, the two
members of (\ref{SR_QCD}) exhibit different $\tau$-behaviour. The
{\em exact\/} effective continuum threshold, which would reproduce
the {\em true\/} values of hadron mass and decay constant on the
left-hand side of (\ref{SR_QCD}), is, clearly, not known.
Therefore, our ideas of {\em extracting\/} hadron parameters from
sum rules consist in attempting to obtain a reliable approximation
to the exact threshold $s_{\rm eff}$ and to control the accuracy
of this approximation. In a recent series of publications
\cite{lms_new}, we have constructed all procedures, techniques,
and algorithms required to achieve this~goal: With our concept of
$s_{\rm eff}(\tau),$ we define dual mass $M_{\rm dual}$ and dual
decay constant $f_{\rm dual}$ of~$P$~by$$M_{\rm
dual}^2(\tau)\equiv-\frac{{\rm d}}{{\rm d}\tau}\log\Pi_{\rm
dual}(\tau,s_{\rm eff}(\tau))\ ,\qquad f_{\rm dual}^2(\tau)\equiv
M_P^{-4}\exp(M_P^2\,\tau)\,\Pi_{\rm dual}(\tau,s_{\rm eff}(\tau))\
.$$If the ground-state mass' actual value $M_P$ is known, the
deviation of our dual ground-state mass $M_{\rm dual}$ from this
$M_P$ indicates the amount of excited-state contributions picked
up by our dual correlator $\Pi_{\rm dual}(\tau,s_{\rm
eff}(\tau)).$ Assuming specific Ans\"atze for our function $s_{\rm
eff}(\tau)$ and requiring least deviation of our $M_{\rm dual}$
from the true $M_P$ in the range of admissible values of the Borel
parameter $\tau$ leads to a variational solution for the effective
threshold. With $s_{\rm eff}(\tau)$ at hand, we find the
$P$-meson's decay constant from the second of the above dual
relations. The traditional {\em assumption\/} for the effective
threshold is that it is a ($\tau$-independent) constant. In
addition to this very crude approximation, we consider for $s_{\rm
eff}(\tau)$ also polynomials in $\tau.$ It is easy to imagine that
a $\tau$-dependent threshold greatly facilitates reproducing the
true mass value $M_P.$ This implies that a dual correlator with
$\tau$-dependent threshold isolates the ground state to much
higher extent and is less plagued by excited-state contamination
than a dual correlator with the conventional, but na\"ive,
$\tau$-independent threshold. Consequently, the accuracies of
extracted hadron observables are drastically improved. Recent
experience from various quantum-mechanical test grounds reveals
that the band of results computed from linear, quadratic, and
cubic Ans\"atze for $s_{\rm eff}(\tau)$ encompasses the exact
$f_P$ value \cite{lms_new} and that the extraction procedures in
quantum mechanics and in QCD are (even quantitatively) very
similar \cite{lms_qcdvsqm}. For all the details of our improved
sum-rule approach, consult Refs.~\cite{lms2010,lms2011,lms_1,
lms_new,lms_qcdvsqm,lms2010_conf}.

{\bf OPE and Heavy-Quark Mass Scheme.} A close inspection shows
that for both heavy-light correlators and resulting decay
constants the choice of the precise mass scheme adopted for
defining the heavy-quark mass is crucial. The OPE for the
correlator (\ref{SR_QCD}) to three-loop order was derived in terms
of the heavy-quark {\em pole mass\/} in \cite{chetyrkin}. An
alternative is to reorganize the perturbative expansions in terms
of the heavy-quark {\em running $\overline{\it MS}$ mass\/}
\cite{jamin}.~Figure~\ref{Plot:1} presents the $B$-meson decay
constant $f_B$ resulting from both choices. In each case, a {\em
constant\/} effective threshold [differing, of course, for pole
($s_0$) and $\overline{\rm MS}$ ($\overline{s_0}$) mass scheme] is
fixed by requiring maximum stability of the $f_B$ value obtained.
From this exercise we gain important insights:

\begin{figure}[!b]\begin{tabular}{cc}
\includegraphics[width=7.106cm]{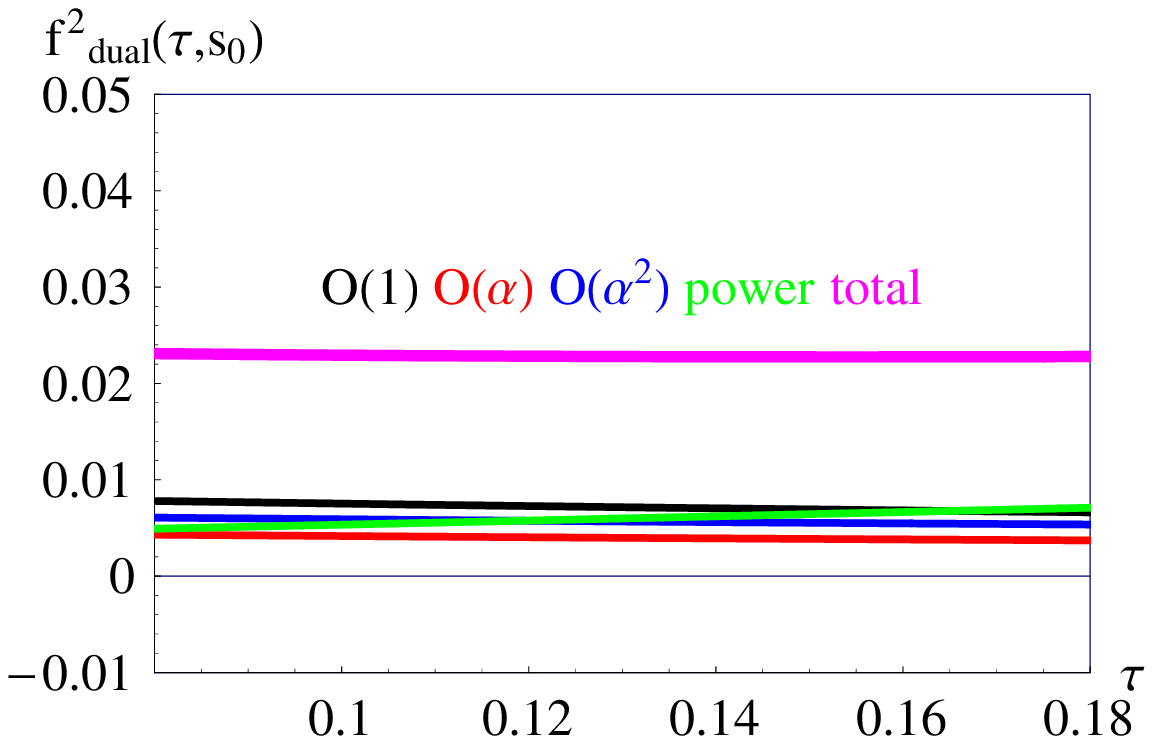}&
\includegraphics[width=7.106cm]{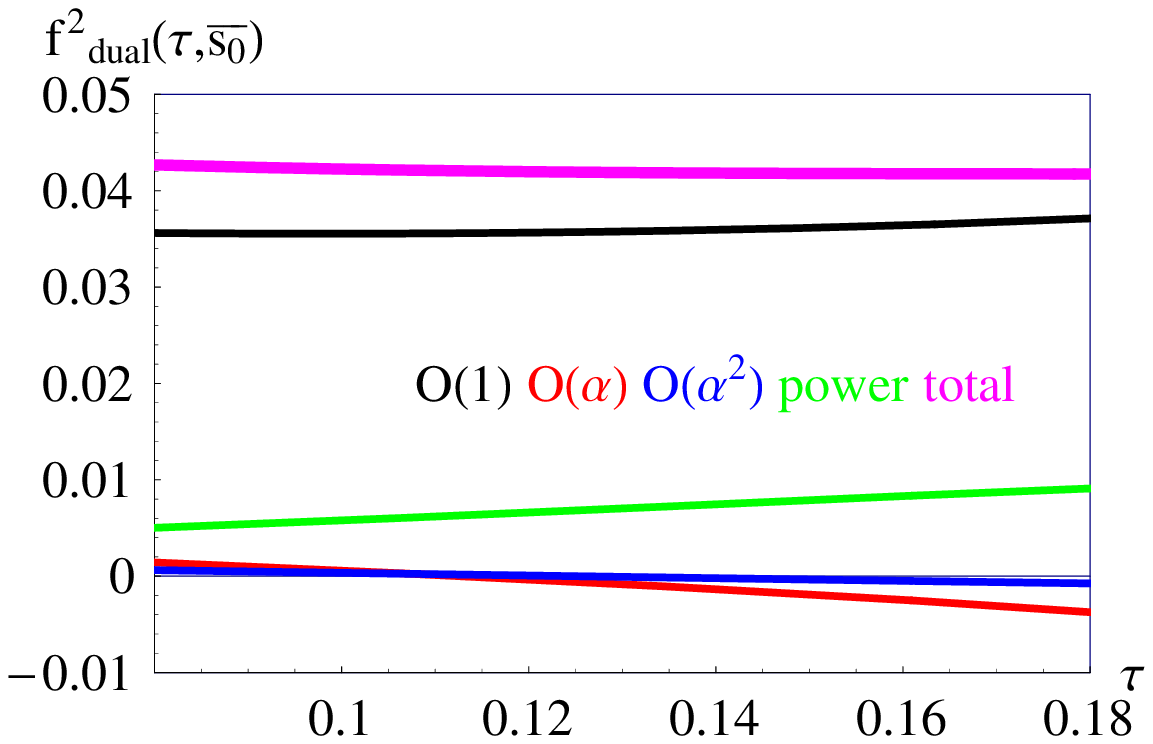}
\end{tabular}\caption{Dual decay constants $f_{\rm dual}$ of the
$B$ meson extracted, for constant thresholds
$\stackrel{\mbox{{\tiny(}{\scriptsize\bf ---}{\tiny )}}}{s_0}$,
from the correlator (\ref{SR_QCD}) expressed in terms of the
$b$-quark's pole (left) and $\overline{\rm MS}$ (right) mass.}
\label{Plot:1}\end{figure}

(a) In the pole-mass scheme, the perturbative series for the decay
constant shows no sign of convergence. The separate contributions
of LO, NLO, and NNLO terms are of similar size. Accordingly, the
pole-mass-scheme result for $f_B$ significantly underestimates its
true~value.

(b) Reorganizing the perturbative series in terms of the
heavy-quark $\overline{\rm MS}$ mass generates an impressively
distinct {\em hierarchy of the perturbative contributions\/}. Our
dual decay constant $f_{\rm dual}$ obtained using the
$\overline{\rm MS}$ scheme proves to be some 40\% larger than in
the pole-mass~scheme.

(c) Interestingly, in {\em both\/} mass schemes the decay constant
exhibits perfect stability in a wide range of the Borel parameter
$\tau$. This clearly tells us that mere Borel stability is not
sufficient to ensure the reliability of a sum-rule extraction of
bound-state features. Repeatedly \cite{lms_1}, we tried to draw
the attention of sum-rule practitioners to this observation;
nevertheless, some authors seem to be content with Borel stability
as a proof of the trustability of their~findings.

In view of the above, we adopt the OPE expressed in terms of
running $\overline{\rm MS}$ quark masses~\cite{jamin}.

{\bf\boldmath Decay Constants of $D$ and $D_s$
\cite{lms2010,lms2011}.} Straightforward application of our
sum-rule algorithm yields, as our predictions for the decay
constants of the charmed pseudoscalar mesons $D_{(s)},$
\begin{align*}f_D&=(206.2\pm7.3_{\rm(OPE)}\pm5.1_{\rm(syst)})\;\mbox{MeV}\
,\\f_{D_s}&=(245.3\pm15.7_{\rm(OPE)}\pm4.5_{\rm(syst)})\;\mbox{MeV}\
.\end{align*}Herein, the OPE-related errors are computed by
bootstrap studies allowing the parameters induced by QCD (i.e.,
quark masses, $\alpha_{\rm s}$, and condensates) to vary in their
respective ranges. We observe perfect agreement of our results
with the corresponding lattice QCD outcomes. Let us emphasize that
the $\tau$-dependent effective threshold constitutes {\em the\/}
crucial ingredient for a successful prediction of decay constants
of charmed heavy mesons by the sum rule (\ref{SR_QCD}). Standard
$\tau$-independent approximations entail a much lower value for
the $D$-meson decay constant, $f_D,$ that resides rather far from
both the experimental data {\em and\/} the lattice~findings.

{\bf\boldmath Decay Constants of $B$ and $B_s$ \cite{lms2010}.}
Our QCD sum-rule results for the decay constants~$f_{B_{(s)}}$~of
the pseudoscalar beauty mesons $B_{(s)}$ turn out to be extremely
sensitive to the input value of the $b$-quark mass; for instance,
the $b$-quark's $\overline{\rm MS}$-mass range
$\overline{m}_b(\overline{m}_b)=(4.163\pm0.016)\;\mbox{GeV}$
\cite{mb} gives results that are barely compatible with recent
lattice computations of these decay constants. However, inverting
the logic by requiring our sum-rule result for $f_B$ to~match~the
average of these lattice calculations provides the very precise
value of the $b$-quark $\overline{\rm MS}$ mass
$$\overline{m}_b(\overline{m}_b)=(4.245\pm0.025)\;{\rm GeV}\
.$$The corresponding estimates for $f_B$ and $f_{B_s}$ emerging
within our sum-rule prescriptions~are\begin{align*}
f_B&=(193.4\pm12.3_{\rm(OPE)}\pm4.3_{\rm(syst)})\;{\rm MeV}\
,\\f_{B_s}&=(232.5\pm18.6_{\rm(OPE)}\pm2.4_{\rm(syst)})\;{\rm
MeV}\ .\end{align*}

{\bf Summary and Conclusions.}

{\bf 1.}~The $\tau$-dependence of effective thresholds emerges
naturally when one attempts to render the duality relation exact.
Let us emphasize two facts: (a) In principle, this
$\tau$-dependence is {\em not\/} in conflict with the properties
of quantum field theories. (b) Our analysis of $D_{(s)}$~mesons
indicates that it will indeed raise the quality of the resulting
sum-rule predictions {\em decisively\/}.

{\bf 2.}~Our study of {\em charmed mesons\/} clearly demonstrates
that using Borel-parameter-dependent thresholds leads to lots of
essential improvements: (i) The accuracy of sum-rule predictions
for decay constants is significantly increased. (ii) It has become
possible to extract a realistic systematic error and to diminish
it to the level of a few percent. (iii) Our prescription brings
the QCD sum-rule approach into perfect agreement with both lattice
QCD and experiment.

{\bf 3.}~The {\em beauty-meson\/} decay constants $f_{B_{(s)}}$
are extremely sensitive to the choice of the $b$-quark mass:
Regarding this as a kind of serendipity and matching our QCD
sum-rule~outcome for $f_B$ to the corresponding average of lattice
evaluations enables us to arrive at a rather precise estimate of
$\overline{m}_b(\overline{m}_b)$ in good agreement with several
lattice results but which, unfortunately, has no overlap with a
recent, rather accurate determination \cite{mb}; for details,
consult Ref.~[2].

\vspace{3ex}\noindent{\em Acknowledgments\/}. DM is supported by
the Austrian Science Fund (FWF), project no.~P22843.

\bibliographystyle{aipproc}

\begin{thebibliography}{99}
\bibitem{svz}M.~A.~Shifman, A.~I.~Vainshtein, and V.~I.~Zakharov,
Nucl.~Phys.~B {\bf 147}, 385 (1979).
\bibitem{lms2010}W.~Lucha, D.~Melikhov, and S.~Simula, J.~Phys.~G:
Nucl.~Part.~Phys.~{\bf 38}, 105002 (2011).
\bibitem{lms2011}W.~Lucha, D.~Melikhov, and S.~Simula,
Phys.~Lett.~B {\bf 701}, 82 (2011); arXiv:1108.0844 [hep-ph].
\bibitem{lms_1}W.~Lucha, D.~Melikhov, and S.~Simula,
Phys.~Rev.~D~{\bf 76}, 036002 (2007); Phys.~Lett.~B~{\bf 657}, 148
(2007); Phys.~Atom.~Nucl.~{\bf 71}, 1461 (2008); Phys.~Lett.~B
{\bf 671}, 445 (2009); arXiv:1107. 1848 [hep-ph]; D.~Melikhov,
Phys.~Lett.~B {\bf 671}, 450 (2009).
\bibitem{lms_new}W.~Lucha, D.~Melikhov, and S.~Simula, Phys.~Rev.~D
{\bf 79}, 096011 (2009); J.~Phys.~G: Nucl.\ Part.~Phys.~{\bf 37},
035003 (2010); W.~Lucha, D.~Melikhov, H.~Sazdjian, and S.~Simula,
Phys.\ Rev.~D {\bf 80}, 114028 (2009).
\bibitem{lms_qcdvsqm}W.~Lucha, D.~Melikhov, and S.~Simula,
Phys.~Lett.~B {\bf 687}, 48 (2010); Phys.~Atom.~Nucl.\ {\bf 73},
1770 (2010); in {\em QCD@Work 2010\/}, eds.~L.~Angelini {\em et
al.}, AIP Conf.~Proc.~{\bf 1317} (AIP, Melville, New York, 2010),
p.~316, arXiv:1008.0167 [hep-ph].
\bibitem{lms2010_conf}W.~Lucha, D.~Melikhov, and S.~Simula, in {\em
QCD@Work 2010\/}, eds.~L.~Angelini {\em et al.}, AIP
Conf.~Proc.~{\bf 1317} (AIP, Melville, New York, 2010), p.~310,
arXiv:1008.3129 [hep-ph];~PoS (ICHEP 2010) 210; PoS (QFTHEP2010)
058.
\bibitem{chetyrkin}K.~G.~Chetyrkin and M.~Steinhauser,
Phys.~Lett.~B {\bf 502}, 104 (2001); Eur.~Phys.~J.~C~{\bf 21},~319
(2001).
\bibitem{jamin}M.~Jamin and B.~O.~Lange, Phys.~Rev.~D {\bf 65},
056005 (2002).
\bibitem{mb}K.~G.~Chetyrkin {\it et al.}, Phys.~Rev.~D {\bf 80},
074010 (2009).\end{thebibliography}
}

\end{document}